\documentclass[12pt]{article}
\usepackage{makeidx}
\usepackage{amsmath}
\usepackage{amsfonts}
\usepackage{amssymb}
\usepackage{epsfig}
\usepackage[cp1251]{inputenc}
\usepackage{graphicx}

\def\bra #1{\langle #1\vert}
\def\ket #1{\vert #1\rangle}

\newcounter{defin}
\newcounter{lemma}
\newcounter{theorem}
\newcounter{proposition}
\newcounter{example}
\newcounter{comment}
\newenvironment{lemma}{\par\refstepcounter{lemma}     \textbf{Lemma \thelemma.} }{\rm\par}
\newenvironment{theorem}{\par\refstepcounter{theorem}     \textbf{Theorem \thetheorem.}\ }{\rm\par}
\newenvironment{proposition}{\par\refstepcounter{proposition}     \textbf{Proposition \theproposition.}\ }{\rm\par}

\begin{document}

\title{Proof of the Gaussian maximizers conjecture for the communication capacity of noisy heterodyne measurements}
\author{A. S. Holevo, S. N. Filippov \\
Steklov Mathematical Institute \\ Russian Academy of Sciences\\
8 Gubkina St., Moscow 119991, Russia}
\date{}
\maketitle

\begin{abstract}
Basing on recently developed convex programming framework in the
paper [arXiv:2204.10626], we provide a proof for a long-standing
conjecture on optimality of Gaussian encondings for the ultimate
communication rate of generalized heterodyne receivers under the
oscillator energy constraint. Our results generalize previous ones
(obtained under the assumption of validity of the energy threshold
condition) and show a drastic difference in the structure of the
optimal encoding within and beyond this condition. The core of the
proof in the case beyond the threshold is a new log-Sobolev type
inequality, which relates the generalized Wehrl entropy with the
wavefunction gradient.
\end{abstract}

\section{Introduction}

Transmission of classical information encoded into quantum states
is of great interest for both fundamental and practical reasons.
Fundamentally, the classical capacity of a quantum channel defines the
ultimate reliable communication
rate~\cite{holevo-1998,schumacher-westmoreland-1997}. In reality,
any physical signal has underlying quantum-mechanical nature
which has to be taken into account. The latter fact is of central
importance also for the measuring apparatus because of the
quantum-mechanical complementarity, which sets up an upper
bound for distinguishability of quantum signals.

In this paper we focus on the particular form of measurement ---
generalized heterodyne detection --- which can be regarded as an
approximate measurement of two non-commuting canonical variables
(e.g., field quadratures playing the role of position and
momentum). In the process of heterodyning, the signal density
operator is mapped to a generalized Husimi
function~\cite{lalovic-1992}. Quantum description of noisy
heterodyne receivers and equivalent devices is given and reviewed,
e.g., in
Refs.~\cite{yuen-shapiro-1980,walker-carroll-1984,shapiro-1985,stenholm-1992,caves,sera}.

The generalized heterodyne measurement itself can be viewed as a specific
quantum-classical Gaussian channel~\cite{acc-noJ,entropy}, whose
capacity is to be determined and the optimal encoding is to be
found. This brings us to a long-standing Gaussian maximizer conjecture that the energy-constrained classical capacity
of a general quantum Gaussian channel is always attained on a Gaussian encoding.
Although many authors have evaluated the communication rate
of Gaussian channels by using Gaussian encodings (e.g., in
\cite{hall3,scha,guha,lee}), the obtained evaluations give only a
lower bound for the classical capacity unless the conjecture is
proved.

The conjecture on optimality of Gaussian encodings was proved
correct for gauge covariant and contravariant bosonic Gaussian
channels (called phase-insensitive in quantum
optics)~\cite{ghg,ghm,ghm1}, and later for a broader class of
channels satisfying ``threshold condition'' under which the upper
bound for the capacity as a difference between the maximum and the
minimum output entropies is attainable~\cite{h2,acc2}.  However,
the conjecture remained open for a variety of other quantum and
quantum-classical channels beyond the scope of the threshold
condition~\cite{entropy} including the class of quantum-classical
channels, such as noisy homodyne and heterodyne detection.
Recently the conjecture was proved for a class of
quantum-classical channels describing noisy homodyne
detection~\cite{holevo-2022}. In the proof of~\cite{holevo-2022},
a generalization of the celebrated log-Sobolev
inequality~\cite{ledoux,lieb} appeared in the context of the
convex optimization problem. In the present paper we make a step
further and extend the proof to noisy heterodyne detection. Here
we prove yet another log-Sobolev type inequality
(Proposition~\ref{p1}), which enables us to extend the treatment
beyond the scope of the threshold condition. Notably, the derived
inequality is also a far-reaching generalization of the Wehrl
inequality proved by Lieb~\cite{lieb1}. Within the threshold
condition we recover the result of Ref.~\cite{acc-noJ} for the
validity of the conjecture in this case. We also show that the
optimal encodings significantly differ within and beyond the
threshold condition, thus clarifying the physical meaning of the
latter.

\section{Capacity of noisy heterodyning: threshold condition}

In this section we briefly summarize the results of
Refs.~\cite{acc-noJ,entropy} concerning the unsharp joint
position-momentum measurement (with the noisy optical heterodyning
as the physical prototype). Statistics of the measurement outcome
$(x,y) \in {\mathbb R}^2$ is described by the following positive
operator-valued measure (POVM):
\begin{equation}
M(dxdy)=D(x,y)\rho _{\beta }D(x,y)^{\ast }\frac{dxdy}{2\pi },
\label{MTB}
\end{equation}%
where $q$ and $p$ are the canonical position and momentum
operators, respectively, $D(x,y)=\exp[ i\left( yq-xp\right)]$ is a
unitary position-momentum displacement operators, and
$\rho_{\beta}$ is a centered Gaussian density operator with the
covariance matrix
\begin{equation}
\beta =\left[
\begin{array}{cc}
\beta _{q} & 0 \\
0 & \beta _{p}%
\end{array}%
\right] ;\quad \beta _{q}\beta _{p}\geq \frac{1}{4}.  \label{beta}
\end{equation}%
Here $\beta _{q}$ ($\beta _{p})$ is the noise power in position
(momentum) quadrature. We denote
\begin{equation}
m(x,y)=\frac{1}{2\pi }D(x,y)\rho _{\beta }D(x,y)^{\ast }.
\label{mxy}
\end{equation}

\noindent For a given system density operator $\rho$ the output
differential entropy\footnote{The differential entropy is well-defined in this case because
the probability density ${\rm Tr} [\rho m(x,y)]$ is uniformly bounded \cite{acc}.} reads
\begin{equation} \label{Wehrl}
h_M(\rho) = - \int {\rm Tr} [\rho m(x,y)] \ln {\rm Tr} [\rho
m(x,y)] \, dx \, dy
\end{equation}

\noindent and represents a generalization of the Wehrl entropy.
Eq.~\eqref{Wehrl} reduces to the conventional Wehrl entropy when
$\beta_{q} = \beta_{p} = \frac{1}{2}$.

Following the lines of Ref.~\cite{holevo-2022}, an encoding ${\cal
E}$ can be viewed as a probability distribution $\pi(d\rho)$ on
the set of quantum states $\mathfrak{S}$. The average state is
\begin{equation}
\bar{\rho}_{\mathcal{E}}=\int_{\mathfrak{S}}\rho \pi (d\rho ).
\end{equation}

The classical Shannon information between the input (encoded into
quantum states $\rho$ with probability distribution $\pi(d\rho)$)
and the measurement outcome $(x,y)$ equals
\begin{equation}
I(\mathcal{E},M) = h_M(\bar{\rho}_{\mathcal{E}}) -
\int_{\mathfrak{S}} h_M(\rho) \pi(d\rho).
\end{equation}

Let $H = \frac{1}{2}(q^2 + p^2)$ be the system Hamiltonian and $E
\in [\frac{1}{2},+\infty)$ be a maximally permissible average
energy. Then the oscillator-energy-constrained classical capacity
of the quantum-classical measurement channel under consideration is
\begin{eqnarray}
C(M,H,E) & = &
\sup_{\mathcal{E}:\mathrm{Tr}\bar{\rho}_{\mathcal{E}}H \leq E} I(
\mathcal{E},M)   \label{capacity} \\
& = & \sup_{\mathcal{E}:\mathrm{Tr}\bar{\rho}_{\mathcal{E}}H \leq
E} \left[ h_M(\bar{\rho}_{\mathcal{E}}) -
e_M(\bar{\rho}_{\mathcal{E}}) \right], \label{capacity-eM}
\end{eqnarray}%

\noindent where the quantity $e_{M}(\rho')$ is defined through
\begin{equation}
e_{M}(\rho') = \inf_{\mathcal{E}:\bar{\rho}_{\mathcal{E}}=\rho'}
\int h_{M}(\rho)\pi(d\rho)
\end{equation}%
and represents an analogue of the convex closure of the output
differential entropy for a quantum channel~\cite{Shir}. Since any
measurement channel is entanglement-breaking, its classical
capacity is additive~\cite{shor-2002,QSCI} and is given by the
single-letter expression~\eqref{capacity}.

An important result of Ref.~\cite{entropy} is that for a general
multi-mode Gaussian measurement channel and a quadratic
Hamiltonian $H$ the supremum in Eqs.~\eqref{capacity},
\eqref{capacity-eM} is attained at a centered Gaussian density
operator $\rho_{\alpha}$ with the covariance matrix $\alpha$. In
the case of a single mode and the oscillator Hamiltonian $H =
\frac{1}{2}(q^2 +
p^2)$ we are dealing with in the present paper, the covariance matrix $\alpha = \left(%
\begin{array}{cc}
  \alpha_q & 0 \\
  0 & \alpha_p \\
\end{array}%
\right)$, and the energy constraint reads as
\begin{equation}
\alpha_q + \alpha_p \leq 2E.
\end{equation}

\noindent Thus, while the  optimal average state
$\bar{\rho}_{\mathcal{E}}$ is available, the structure of the encoding ${\cal
E}$ itself is not known in general. \emph{The Gaussian maximizer
conjecture} states that the optimal encoding consists of squeezed
coherent states with the displacement parameter having a Gaussian
probability distribution. The conjecture was proved to be correct
(in a much more general multi-mode situation) in \cite{acc-noJ}
under the threshold condition which has a form of inequality between the matrix $\alpha$ and
the covariance matrix of the squeezed state minimizing the output entropy. In the single-mode case
it takes the form (see \cite{entropy}):

\textbf{Case C (Central).} The threshold condition on the parameters $\alpha
_{q}$, $\alpha _{p}$, $\beta _{q}$, $\beta _{p}$ reduces to the inequalities:%
\begin{equation}
\frac{1}{2\alpha _{p}}<\sqrt{\frac{\beta _{q}}{\beta
_{p}}}<2\alpha _{q}. \label{thr}
\end{equation}

Assuming this condition to hold, the optimal encoding is Gaussian $\mathcal{E}_0=\left\{ \pi
_{0}(dxdy),\rho _{0}(x,y)\right\} $, where $\rho
_{0}(x,y)=\left\vert x,y\right\rangle _{\delta }\left\langle
x,y\right\vert $ is a squeezed coherent state with the vector
$\left\vert x,y\right\rangle _{\delta }=D(x,y)\left\vert
0\right\rangle _{\delta }$,
\begin{equation}
\delta =\frac{1}{2}\sqrt{\frac{\beta _{q}}{\beta _{p}}}
\label{delta}
\end{equation}%
and $\pi _{0}(dxdy)$ is the centered normal distribution with the
nondegenerate covariance matrix
\begin{equation*}
\gamma =\left[
\begin{array}{cc}
\gamma _{q} & 0 \\
0 & \gamma _{p}%
\end{array}%
\right] =\left[
\begin{array}{cc}
\alpha _{q}-\delta  & 0 \\
0 & \alpha _{p}-1/\left( 4\delta \right)
\end{array}%
\right] .
\end{equation*}%
The communication rate
\begin{equation*}
I({\cal E}_0,M) = \frac{1}{2}\log \frac{\left( \alpha _{q}+\beta
_{q}\right)
\left( \alpha _{p}+\beta _{p}\right) }{\left( \sqrt{\beta _{q}\beta _{p}}%
+1/2\right) ^{2}}.
\end{equation*}%
An additional optimization over $\alpha _{q},\alpha _{p}$
satisfying $\alpha
_{q}+\alpha _{p}\leq 2E$ gives the energy-constrained capacity%
\begin{equation*}
C(M;H,E)=\log \left( \frac{E+\left( \beta _{q}+\beta _{p}\right) /2}{\sqrt{%
\beta _{q}\beta _{p}}+1/2}\right)
\end{equation*}%
for $E$ satisfying the energy threshold%
\begin{equation*}
E\geq \max \left\{ E\left( \beta _{p},\beta _{q}\right) ,E\left(
\beta
_{q},\beta _{p}\right) \right\} ,\quad E\left( \beta _{1},\beta _{2}\right) =%
\frac{1}{2}\left( \beta _{1}-\beta _{2}+\sqrt{\frac{\beta _{1}}{\beta _{2}}}%
\right) .
\end{equation*}%

When the threshold condition is violated, then the Gaussian
maximizer conjecture remained an open problem so far. To develop a
general theory applicable both within and beyond the threshold
condition we exploit a recently introduced framework of
Ref.~\cite{holevo-2022}, which reformulates the optimization
problem in terms of the convex programming.

\section{Problem formulation in terms of convex programming}
\label{section-convex-prog}

Following Ref.~\cite{holevo-2022}, we introduce the functional
\begin{equation}
F(\mathcal{E}) = \int_{\mathfrak{S}} h_M (\rho) \pi (d\rho) =
\int_{\mathfrak{S}}\mathrm{Tr} \left[ K(\rho) \rho \right]
\pi(d\rho),
\end{equation}

\noindent where
\begin{equation} \label{K-definition}
K(\rho) = - \int m(x,y) \ln {\rm Tr} [\rho m(x,y)] \, dx \, dy.
\end{equation}%

\noindent For a fixed state $\rho'$ the calculation of
$e_M(\rho')$ reduces to the optimization problem over
distributions of density operators:
\begin{eqnarray}
\text{Minimize~} \int_{\mathfrak{S}}\mathrm{Tr} \left[ K(\rho)
\rho \right] \pi(d\rho) \text{~~subject~to~} \int \rho \,
\pi(d\rho) = \rho',
\end{eqnarray}

\noindent which is formally similar to a general Bayes problem
studied in Refs.~\cite{jma1,hol}. Ref.~\cite{holevo-2022} provides
the following conditions for optimality of an encoding
$\mathcal{E} _{0}$ with the distribution $\pi_0(d\rho)$: There
exists a Hermitian operator $\Lambda _{0}$ such that

\begin{itemize}

\item[(i)] $\Lambda_{0}\leq K(\rho )$ for all $\rho \in
\mathfrak{S}$;

\item[(ii)] $\left[ K(\rho) - \Lambda_{0}\right] \rho = 0$ almost
everywhere with respect to $\pi_0(d\rho)$.

\end{itemize}

In Refs.~\cite{jma1,hol} mathematical theorems were proved giving precise
regularity assumptions under which the conditions (i), (ii) are necessary and sufficient for the optimality.
In solving our
capacity problem for the Gaussian measurements we will use these conditions as sufficient in a broader context
involving unbounded bosonic operators (see \cite{holevo-2022} for a justification).

By integrating (ii), we get an equation for determination of
$\Lambda_{0}$, namely,
\begin{equation}
\int_{\mathfrak{S}} K(\rho) \, \rho \, \pi_{0}(d\rho) =
\Lambda_{0} \rho'.  \label{iii}
\end{equation}%
However, a major difficulty may be
to check the operator inequality (i).

Let us illustrate this in the case C (within the threshold condition) for noisy
heterodyning (for which there is an alternative proof of optimality \cite{acc-noJ}).
A candidate for the optimal encoding can be guessed within the
class of Gaussian encodings. We will prove the optimality conditions (i) and (ii) for
the Gaussian encoding $\mathcal{E}_{0}=\left\{ \pi _{0}(dxdy),\rho
_{0}(x,y)\right\}$, where
\begin{equation}
\rho _{0}(x,y)=\left\vert x,y\right\rangle_{\delta} \!
\left\langle x,y\right\vert = D(x,y)\left\vert 0\right\rangle
_{\delta } \! \left\langle 0\right\vert D(x,y)^{\ast }.
\label{khet}
\end{equation}%
We start with computation of $K(\rho _{0}(x,y))\rho _{0}(x,y)$ for
an arbitrary $\delta$ and then focus on the case C where $\delta =\frac{1}{2}\sqrt{\frac{%
\beta _{q}}{\beta _{p}}}$.

Using Eq.~\eqref{K-definition} and the explicit formula
\begin{equation}
_{\delta} \langle x',y' \vert m(x,y) \vert x',y' \rangle_{\delta}
= \frac{\exp\left( - \dfrac{(x'-x)^2}{2( \beta _{q}+\delta )} -
\dfrac{(y'-y)^2}{2( \beta_{p}+1/4\delta )} \right)}{2\pi \sqrt{
\left( \beta _{q}+\delta \right) \left( \beta _{p}+1/4\delta
\right)}}
\end{equation}

\noindent as well as introducing $c=\ln 2\pi \sqrt{\left( \beta
_{q}+\delta \right) \left( \beta _{p}+1/4\delta \right) }$, we get
\begin{eqnarray*}
K(\rho _{0}(x^{\prime },y^{\prime })) &=& \int D(x,y)\rho _{\beta
}D(x,y)^{\ast } \left[ c+\frac{\left( x^{\prime }-x\right)
^{2}}{2\left( \beta _{q}+\delta \right) }+\frac{\left( y^{\prime
}-y\right) ^{2}}{2\left( \beta
_{p}+1/4\delta \right) }\right] \frac{dx \, dy}{2\pi } \\
&=&c+\frac{\left( q-x^{\prime }\right) ^{2}+\beta _{q}}{2\left(
\beta
_{q}+\delta \right) }+\frac{\left( p-y^{\prime }\right) ^{2}+\beta _{p}}{%
2\left( \beta _{p}+1/4\delta \right) }.
\end{eqnarray*}%

\noindent Here we also used the formulas (see, e.g., \cite{asp})
\begin{eqnarray}
&& \int D(x,y) \rho _{\beta } D(x,y)^{\ast }\frac{dx \, dy}{2\pi
}=I, \label{int-I} \\
&& \int x^{2}D(x,y)\rho _{\beta }D(x,y)^{\ast }\frac{dx \, dy}{2\pi }%
=q^{2}+\beta _{q}, \label{int-x2} \\
&& \int y^{2}D(x,y)\rho _{\beta }D(x,y)^{\ast }\frac{dx \, dy}{2\pi }%
=p^{2}+\beta _{p}. \label{int-y2}
\end{eqnarray}%

\noindent Hence
\begin{eqnarray}
K(\rho _{0}(x,y))\rho _{0}(x,y) &=&\left[ c+\frac{\left(
q-x\right) ^{2}+\beta _{q}}{2\left( \beta _{q}+\delta \right)
}+\frac{\left( p-y\right) ^{2}+\beta _{p}}{2\left( \beta
_{p}+1/4\delta \right) }\right] D(x,y)\left\vert 0\right\rangle
_{\delta }\left\langle x,y\right\vert  \notag
\\
&=&D(x,y)\left[ c+\frac{q^{2}+\beta _{q}}{2\left( \beta _{q}+\delta \right) }%
+\frac{p^{2}+\beta _{p}}{2\left( \beta _{p}+1/4\delta \right)
}\right]
\left\vert 0\right\rangle _{\delta }\left\langle x,y\right\vert   \notag \\
&=&\left[ c+\frac{\beta _{q}}{2\left( \beta _{q}+\delta \right) }+\frac{%
\beta _{p}}{2\left( \beta _{p}+1/4\delta \right) }\right]
\left\vert
x,y\right\rangle _{\delta }\left\langle x,y\right\vert  \notag \\
&&+D(x,y)\frac{1}{2}\left[ \frac{q^{2}}{ \beta _{q}+\delta  }+%
\frac{p^{2}}{ \beta _{p}+1/4\delta  }\right] \left\vert
0\right\rangle _{\delta }\left\langle x,y\right\vert .
\label{Khet}
\end{eqnarray}%

In the case C, we consider $\delta$ given by Eq.~\eqref{delta}.
For this value of $\delta $,
\begin{equation*}
\frac{\beta _{q}+\delta  }{ \beta _{p}+1/4\delta }=\frac{2\delta
}{1/2\delta },
\end{equation*}%
hence $\left\vert 0\right\rangle _{\delta }$ is the ground state
of the Hamiltonian
\begin{equation*}
\frac{q^{2}}{ \beta _{q}+\delta  }+\frac{p^{2}}{ \beta
_{p}+1/4\delta  }=\frac{2\delta }{\beta _{q}+\delta }\left( \frac{%
q^{2}}{2\delta }+2\delta p^{2}\right)
\end{equation*}%
with the minimal eigenvalue%
\begin{equation*}
\Big[ \left( \beta _{q}+\delta \right) \left( \beta _{p}+1/4\delta
\right) \Big]^{-1/2} = \left( \sqrt{\beta _{q}\beta
_{p}}+1/2\right) ^{-1}.
\end{equation*}

\noindent Substituting the value (\ref{delta}) into
Eq.~\eqref{Khet}, we finally obtain
\begin{eqnarray*}
K(\rho _{0}(x,y))\rho _{0}(x,y) &=&\left[ c + \frac{\sqrt{\beta _{q}\beta _{p}}%
}{ \sqrt{\beta _{q}\beta _{p}}+1/2 } + \frac{1}{2\left( \sqrt{%
\beta _{q}\beta _{p}}+1/2\right) }\right] \left\vert
x,y\right\rangle
_{\delta }\left\langle x,y\right\vert \\
&=&\left[ \ln 2\pi \left( \sqrt{\beta _{q}\beta _{p}}+1/2\right)
+1\right]
\left\vert x,y\right\rangle _{\delta }\left\langle x,y\right\vert \\
&=&\ln 2\pi e\left( \sqrt{\beta _{q}\beta _{p}}+1/2\right)
\left\vert x,y\right\rangle _{\delta }\left\langle x,y\right\vert
.
\end{eqnarray*}%

\noindent Integrating with respect to the probability distribution
$\pi _{0}(dxdy),$ we obtain \eqref{iii} with the Hermitian operator
\begin{equation*}
\Lambda _{0}=\ln 2\pi e\left( \sqrt{\beta _{q}\beta
_{p}}+1/2\right) I.
\end{equation*}

To check the condition (i) it is sufficient to prove
\begin{equation}
\left\langle \psi \right\vert \Lambda _{0}\left\vert \psi
\right\rangle \leq \left\langle \psi \right\vert K(\rho
)\left\vert \psi \right\rangle \label{111b}
\end{equation}%
for an arbitrary density operator $\rho $ and a dense subset of
$\psi$ in the system Hilbert space. We can assume that $\psi $ is
a unit vector. Due to nonnegativity of the classical relative
entropy of two probability densities (the Kullback-Leibler
divergence) we have
\begin{eqnarray*}
\left\langle \psi \right\vert K(\rho )\left\vert \psi
\right\rangle &=& -\int \left\langle \psi \right\vert
m(x,y)\left\vert \psi \right\rangle
\ln \mathrm{Tr} [\rho m(x,y)] \, dx\, dy \\
& = & -\int \left\langle \psi \right\vert m(x,y)\left\vert \psi
\right\rangle \ln \frac{\mathrm{Tr} [\rho m(x,y)]}{\left\langle
\psi \right\vert m(x,y)\left\vert \psi
\right\rangle} \, dx\, dy \\
&& -\int \left\langle \psi \right\vert m(x,y)\left\vert \psi
\right\rangle \ln \left\langle \psi \right\vert m(x,y)\left\vert
\psi \right\rangle dx \, dy \\
&\geq & -\int \left\langle \psi \right\vert m(x,y)\left\vert \psi
\right\rangle \ln \left\langle \psi \right\vert m(x,y) \left\vert
\psi \right\rangle dx \, dy \\
&=& h_{M}(\left\vert \psi \right\rangle \left\langle \psi
\right\vert ),
\end{eqnarray*}%
\noindent where $h_{M}(\left\vert \psi \right\rangle \left\langle
\psi \right\vert )$ is the output differential entropy for the
considered measurement channel (the generalized Wehrl entropy),
which is bounded from below by $\left\langle \psi \right\vert
\Lambda _{0}\left\vert \psi \right\rangle = \ln 2\pi e\left(
\sqrt{\beta _{q}\beta _{p}}+1/2\right)$ due to the following
result (that completes the proof of inequality~\eqref{111b}).

\begin{proposition}\label{p0}
\begin{equation}
\min_{\left\Vert \psi \right\Vert =1}h_{M}(\left\vert \psi \right\rangle
\left\langle \psi \right\vert )=\ln 2\pi e\left( \sqrt{\beta _{q}\beta _{p}}%
+1/2\right) .  \label{minh}
\end{equation}%
\end{proposition}
\noindent Let us outline the proof since it seems was not explained sufficiently
before. Proposition 1 in \cite{ghm}, applied to the entropy function,
implies that the output differential entropy of the measurement channel is
minimized by coherent states (a generalization of Wehrl's inequality). This
proof is for the gauge-invariant case i.e. for the standard complex
structure of multiplication by $i$, but since all the complex structures are
isomorphic via a symplectic conjugation, the result is valid for any complex
structure with the correspondingly squeezed states as the minimizers. The
result (\ref{minh}) which can be found in Sec. IV of the
paper \cite{acc2} (see also Example in \cite{acc-noJ}) corresponds to the
following complex structure
\begin{equation*}
J=\left[
\begin{array}{cc}
0 & -\sqrt{\frac{\beta _{p}}{\beta _{q}}} \\
\sqrt{\frac{\beta _{q}}{\beta _{p}}} & 0%
\end{array}%
\right] .
\end{equation*}%
in the one-mode case. The Wehrl inequality proved by
Lieb~\cite{lieb1} corresponds to $\beta _{q}=\beta _{p}=1/2$ and
gives the minimal entropy $\ln 2\pi e$ attained at the coherent
states.

\section{Capacity of noisy heterodyning: beyond threshold condition}

The threshold condition (\ref{thr}) can be violated in two ways:

\begin{itemize}

\item \textbf{Case L (Left)}. $\frac{1}{2\alpha _{p}}\geq
\sqrt{\frac{\beta _{q}}{\beta _{p}}}$.

\item \textbf{Case R (Right)}. $\sqrt{\frac{\beta _{q}}{\beta
_{p}}}\geq 2\alpha _{q}$.

\end{itemize}

\noindent Since these cases are completely symmetric with respect
to exchange between $q$ and $p$, we will restrict ourselves to the
case L.

The Gaussian maximizer conjecture for the noisy heterodyning in
the case L is formulated in Ref.~\cite{entropy}. Suppose
$\rho_{\alpha}$ is the average state in the optimal encoding, then
the conjecture states that
the optimal encoding itself is Gaussian and takes the form $%
\mathcal{E}_{0}=\left\{ \pi _{0}(dx),\rho _{0}(x)\right\} $, where
\begin{equation}
\pi _{0}(dx)=\frac{1}{\sqrt{2\pi \gamma }}\exp \left[ -\frac{x^{2}}{2\gamma }%
\right] dx,\qquad \rho \,_{0}(x)=\left\vert x\right\rangle
_{\delta }\left\langle x\right\vert
\label{optimal-states-beyond-threshold}
\end{equation}%
with%
\begin{equation*}
\delta =\frac{1}{4\alpha _{p}},\quad \gamma =\alpha
_{q}-\frac{1}{4\alpha _{p}}.
\end{equation*}

\noindent This encoding is the same as in the case of unsharp position
measurement (noisy homodyning in quantum optics) discussed in
Ref.~\cite{holevo-2022}. If the conjecture is correct, then the
oscillator-energy-constrained capacity reads
\begin{equation}  \label{cgaus}
C(M,H,E) = \log \left( \frac{\sqrt{1+8E\beta_q
+4\beta_q^{2}}-1}{2\beta_q} \right)
\end{equation}

\noindent provided $\beta _{q}\leq \beta _{p}$ and $E<E\left(
\beta _{p},\beta _{q}\right)$.

\begin{theorem} The Gaussian encoding described above is optimal
for the constrained classical capacity of generalized heterodyning
in the case L. \label{theorem}
\end{theorem}

We will prove the theorem by using the formalism developed in
Sec.~\ref{section-convex-prog}. To check the condition (ii), we
can begin with the computation of $\Lambda _{0}$ as in the Case C
and then set $y=0$ in Eq.~\eqref{khet} so that $D(x,0)=D(x)$. We
then obtain
\begin{equation*}
K(\rho _{0}(x))=c+\frac{\left( q-x\right) ^{2}+\beta _{q}}{2\left(
\beta _{q}+\delta \right) }+\frac{p^{2}+\beta _{p}}{2\left( \beta
_{p}+1/4\delta \right) }
\end{equation*}%
and%
\begin{eqnarray*}
K(\rho _{0}(x))\rho _{0}(x) &=&\left[ c+\frac{\beta _{q}}{2\left(
\beta _{q}+\delta \right) }+\frac{\beta _{p}}{2\left( \beta
_{p}+1/4\delta \right)
}\right] \left\vert x\right\rangle _{\delta }\left\langle x\right\vert \\
&&+D(x)\frac{\delta }{ \beta _{q}+\delta  }\left( \frac{q^{2}}{%
2\delta }+2\delta p^{2}\right) \left\vert 0\right\rangle _{\delta
}\left\langle x\right\vert \\
&&-D(x)\frac{1}{2}\left[ \frac{4\delta ^{2}}{ \beta _{q}+\delta
}-\frac{1}{\beta _{p}+1/4\delta }\right] p^{2} \left\vert
0\right\rangle _{\delta }\left\langle x\right\vert .
\end{eqnarray*}

\noindent Taking into account that the squeezed vacuum $\left\vert
0\right\rangle _{\delta }$ is the ground state of the
corresponding oscillator Hamiltonian, i.e.,
\begin{equation*}
\left( \frac{q^{2}}{2\delta }+2\delta p^{2}\right) \left\vert
0\right\rangle _{\delta }=\left\vert 0\right\rangle _{\delta },
\end{equation*}%
and $D(x)$ commutes with $p^{2},$ we finally get %
\begin{equation*}
K(\rho _{0}(x)) \rho _{0}(x)=\Lambda _{0} \rho _{0}(x),
\end{equation*}%
where%
\begin{eqnarray}
\Lambda _{0} &=&\ln 2\pi \sqrt{\left( \beta _{q}+\delta \right)
\left( \beta _{p}+1/4\delta \right) }+\frac{\beta _{q}+2\delta
}{2\left( \beta _{q}+\delta \right) }+\frac{\beta _{p}}{2\left(
\beta _{p}+1/4\delta \right)
} \nonumber\\
&&-\frac{1}{2}\left[ \frac{4\delta ^{2}}{ \beta _{q}+\delta }-%
\frac{1}{ \beta _{p}+1/4\delta }\right] \,p^{2} . \label{lambda0}
\end{eqnarray}%
Note that%
\begin{equation*}
\frac{4\delta ^{2}}{ \beta _{q}+\delta }-\frac{1}{ \beta
_{p}+1/4\delta }=\frac{4\delta ^{2}\beta _{p}-\beta _{q}}{\left(
\beta _{q}+\delta \right) \left( \beta _{p}+1/4\delta \right)
}\geq 0
\end{equation*}%
for $\delta \geq \frac{1}{2}\sqrt{\frac{\beta _{q}}{\beta
_{p}}}.$\bigskip

To check the condition (i) it is sufficient to prove
\begin{equation}
\left\langle \psi \right\vert \Lambda _{0}\left\vert \psi
\right\rangle \leq \left\langle \psi \right\vert K(\rho
)\left\vert \psi \right\rangle \label{11b}
\end{equation}%
for an arbitrary density operator $\rho $ and a dense subset of
$\psi$ in the system Hilbert space. Arguing as in the
Sec.~\ref{section-convex-prog}, the inequality (\ref{11b}) will
follow if we prove
\begin{equation}
\left\langle \psi \right\vert \Lambda _{0}\left\vert \psi
\right\rangle \leq -\int \left\langle \psi \right\vert
m(x,y)\left\vert \psi \right\rangle \ln \left\langle \psi
\right\vert m(x,y)\left\vert \psi \right\rangle dx \, dy
\label{wl}
\end{equation}%
for a unit vector $\psi$. With $\Lambda _{0}$ given by
\eqref{lambda0} it amounts to
\begin{equation*}
\int \left\langle \psi \right\vert m(x,y)\left\vert \psi
\right\rangle \ln \left\langle \psi \right\vert m(x,y)\left\vert
\psi \right\rangle dxdy
\end{equation*}%
\begin{equation*}
+\ln 2\pi \sqrt{\left( \beta _{q}+\delta \right) \left( \beta
_{p}+1/4\delta
\right) }+\frac{\beta _{q}+2\delta }{2\left( \beta _{q}+\delta \right) }+%
\frac{\beta _{p}}{2\left( \beta _{p}+1/4\delta \right) }
\end{equation*}%
\begin{equation}
\leq \frac{4\delta ^{2}\beta _{p}-\beta _{q}}{2\left( \beta
_{q}+\delta \right) \left( \beta _{p}+1/4\delta \right) }\int
\left\vert \psi \prime (x)\right\vert ^{2}dx,
\label{log-Sobolev-type-2}
\end{equation}

\noindent where $\psi \prime (x)$ is the derivative of $\psi(x)$,
the wavefunction in position representation.

Note that for $\delta = \frac{1}{2}\sqrt{\frac{\beta _{q}}{\beta
_{p}}}$ the right-hand side vanishes and the inequality turns into
the valid inequality (\ref{minh}) for the case C. However, if
$\delta > \frac{1}{2}\sqrt{\frac{\beta _{q}}{\beta _{p}}}$, then
the inequality~\eqref{log-Sobolev-type-2}
represents a new type of log-Sobolev inequalities as compared to the
known in literature~\cite{lieb,holevo-2022}. It relates the generalized Wehrl entropy
$h_M(\ket{\psi}\bra{\psi})$ with the wavefunction gradient. We present a proof of the
inequality~\eqref{log-Sobolev-type-2} in the next section. This
proof completes the proof of the of the Gaussian maximizer
conjecture in the case L. The results can be readily extended to
the case R by a proper change of variables.

\section{Proof of the log-Sobolev type inequality \eqref{log-Sobolev-type-2}}

To simplify the notation denote $p_{\rho}^{\beta_q \beta_p}(x,y)
:= {\rm Tr} [\rho m(x,y)]$, where $m(x,y)$ is defined through the
Gaussian state $\rho_\beta$ with the covariance matrix $\left(%
\begin{array}{cc}
  \beta_q & 0 \\
  0 & \beta_p \\
\end{array}%
\right)$ via formula~\eqref{mxy}. By $h(P || Q)$ we denote the
the classical relative entropy (Kullback–Leibler divergence of a probability density $P$ from
a probability density $Q$).

\begin{lemma} \label{lemma-KL} The following equality takes place
\begin{eqnarray}
h \left( p_{\rho}^{\beta_q \beta_p} \Big\vert \Big\vert
p_{\ket{0}_\delta \bra{0}}^{\beta_q \beta_p} \right) & = & \int
{\rm Tr} [\rho m(x,y)] \ln
{\rm Tr} [\rho m(x,y)] \, dx \, dy \nonumber\\
&& + \ln 2\pi \sqrt{(\beta_q + \delta) \left(\beta_p +
\frac{1}{4\delta} \right)} \nonumber\\
&& + \frac{\beta_q + \langle q^2 \rangle_{\rho}}{2(\beta_q +
\delta)} + \frac{\beta_p + \langle p^2 \rangle_{\rho}}{2(\beta_p +
\frac{1}{4\delta})}, \label{KL-heterodyne}
\end{eqnarray}

\noindent where $\langle q^2 \rangle_{\rho} = {\rm Tr}[\rho q^2]$
and $\langle p^2 \rangle_{\rho} = {\rm Tr}[\rho p^2]$.

\end{lemma}

\textit{Proof}. Using the explicit form
\begin{equation}
p_{\ket{0}_\delta \bra{0}}^{\beta_q \beta_p}(x,y) =
\frac{\exp\left( -\dfrac{x^2}{2(\beta_q + \delta)}
-\dfrac{y^2}{2(\beta_p + \frac{1}{4\delta})} \right)}{2\pi
\sqrt{(\beta_q + \delta)(\beta_p + \frac{1}{4\delta})}}
\end{equation}

\noindent and properties \eqref{int-I}, \eqref{int-x2},
\eqref{int-y2}, we readily get Eq.~\eqref{KL-heterodyne} by
the definition of the relative entropy. \hfill $\square$

\begin{proposition}\label{p1}
For all density operators $\rho$ and real positive parameters
$\beta_q,\beta_p,\delta$ satisfying $\beta_q \beta_p \geq
\frac{1}{4}$ and $\delta \geq \frac{1}{2}
\sqrt{\frac{\beta_q}{\beta_p}}$ the following inequality holds:
\begin{eqnarray}
&& \int {\rm Tr} [\rho m(x,y)] \ln
{\rm Tr} [\rho m(x,y)] \, dx \, dy \nonumber\\
&& + \ln 2\pi \sqrt{(\beta_q + \delta)\left(\beta_p +
\frac{1}{4\delta} \right)} + \frac{\beta_q + 2\delta}{2(\beta_q +
\delta)} + \frac{\beta_p}{2(\beta_p + \frac{1}{4\delta})}
\nonumber\\
&& \leq \frac{4 \delta^2 \beta_p - \beta_q}{2(\beta_q +
\delta)(\beta_p + \frac{1}{4\delta})} \langle p^2 \rangle_{\rho}.
\label{het-inequality}
\end{eqnarray}
\end{proposition}

\textit{Proof}. Using Lemma~\ref{lemma-KL}, we rewrite the desired
inequality \eqref{het-inequality} in the form
\begin{equation} \label{KL-inequality}
h \left( p_{\rho}^{\beta_q \beta_p} \Big\vert \Big\vert
p_{\ket{0}_\delta \bra{0}}^{\beta_q \beta_p} \right) \leq
\frac{\langle q^2 \rangle_{\rho} + 4 \delta^2 \langle p^2
\rangle_{\rho} - 2\delta}{2(\beta_q + \delta)}.
\end{equation}

For arbitrarily fixed $\beta_q > 0$ and $\delta > 0$ define
$\widetilde{\beta}_p := \frac{\beta_q}{4 \delta^2}$. To prove the
proposition we need to prove the validity of \eqref{KL-inequality}
for all $\beta_p \geq \widetilde{\beta}_p$.

If $\beta_p = \widetilde{\beta}_p$, then the original inequality
\eqref{het-inequality} reduces to
\begin{equation}
\int {\rm Tr} [\rho m(x,y)] \ln {\rm Tr} [\rho m(x,y)] \, dx \, dy
+ \ln 2\pi (\sqrt{\beta_q \beta_p} + \tfrac{1}{2}) + 1 \leq 0,
\label{inequality-known}
\end{equation}

\noindent which was proved to be correct in the earlier
work~\cite{acc2} (see also Proposition~\ref{p0} in
Sec.~\ref{section-convex-prog}). Therefore, the equivalent
inequality \eqref{KL-inequality} holds true if $\beta_p =
\widetilde{\beta}_p$.

If $\beta_p > \widetilde{\beta}_p$, then $p_{\rho}^{\beta_q
\beta_p} = T_{\beta_p - \widetilde{\beta}_p} p_{\rho}^{\beta_q
\widetilde{\beta}_p}$, where the Markov operator $T_{t}$ acts on a
probability density $P(x,y)$ as follows:
\begin{equation}
T_{t} P(x,y) = \frac{1}{\sqrt{2\pi t}} \int \exp\left( -
\frac{(y-v)^2}{2t} \right) P(x,v) \, dv.
\end{equation}

\noindent Due to monotonicity of the classical relative entropy under Markov operators (classical channels), we have
\begin{eqnarray}
h \left( p_{\rho}^{\beta_q \beta_p} \Big\vert \Big\vert
p_{\ket{0}_\delta \bra{0}}^{\beta_q \beta_p} \right) &=& h \left(
T_{\beta_p - \widetilde{\beta}_p} p_{\rho}^{\beta_q
\widetilde{\beta}_p} \Big\vert \Big\vert T_{\beta_p -
\widetilde{\beta}_p} p_{\ket{0}_\delta
\bra{0}}^{\beta_q \widetilde{\beta}_p} \right) \nonumber\\
& \leq & h \left( p_{\rho}^{\beta_q \widetilde{\beta}_p} \Big\vert
\Big\vert p_{\ket{0}_\delta \bra{0}}^{\beta_q \widetilde{\beta}_p}
\right)
\nonumber\\
&\leq & \frac{\langle q^2
\rangle_{\rho} + 4 \delta^2 \langle p^2 \rangle_{\rho} -
2\delta}{2(\beta_q + \delta)},
\end{eqnarray}

\noindent where the last inequality is just the same as (\ref{inequality-known}). \hfill $\square$

Note that equality in \eqref{KL-inequality} and, consequently, in
\eqref{het-inequality} takes place if $\rho =
\ket{0}_{\delta}\bra{0}$ because both parts of
\eqref{KL-inequality} are equal to zero in this case. Substituting
$\rho = \ket{\psi}\bra{\psi}$ for $\rho$
in~\eqref{het-inequality}, we get the log-Sobolev type
inequality~\eqref{log-Sobolev-type-2} and, hence, the optimality
of the Gaussian encoding in the case L. This concludes the proof
of Theorem~\ref{theorem}.

\section{Conclusions}

We have resolved the Gaussian maximizer conjecture for noisy
heterodyne measurement channel in the affirmative. We treated both
threshold scenarios (within the threshold and beyond it) on an
equal footing by using the recently developed reformulation of the
problem in terms of the convex programming~\cite{holevo-2022}.
This allowed us to reproduce the known results within the
threshold and achieve new results beyond the threshold. The
optimal encodings to attain the oscillator-energy-constrained
capacity are shown to be Gaussian in both scenarios, however, the
structure of the optimal encoding is different: one should use a
two-parameter family of states \eqref{khet} within the threshold
condition and a one-parameter family of states
\eqref{optimal-states-beyond-threshold} beyond the threshold.

Physical meaning of such a structure is clear: when the ratio of
noises $\beta_p/\beta_q$ is  greater than certain threshold value,
the optimal choice for the encoding becomes to invest all the data
into position $q$, leaving momentum $p$ ignored. The optimal
encoding is then the same as in the homodyne case
\cite{holevo-2022} which can be regarded as a limiting case
$\beta_p\rightarrow +\infty$.

Note that beyond the threshold $e_M(\bar{\rho}_{{\cal E}_0}) \neq
\min_{\rho }h_{M}(\rho)$, which means that the previously
known methods (e.g., those used in Refs.~\cite{ghg,ghm,ghm1}) are
not applicable to this case. The conjecture was finally resolved
by proving a new inequality~\eqref{log-Sobolev-type-2} that
relates the generalized Wehrl entropy with the wavefunction
gradient. The new inequality can be considered as another
generalization of the log-Sobolev inequality in addition to the
recently proposed in Ref.~\cite{holevo-2022}.

\section*{Acknowledgment} The work was supported by the grant of Russian Scientific Foundation
19-11-00086, https://rscf.ru/project/19-11-00086/.

\end{document}